\documentclass[]{aastex631}

\accepted{November 26, 2024}

\begin{document}

\title{Evidence for Mass-dependent Evolution of Transitional Dwarf Galaxies in the Virgo Cluster}

\correspondingauthor{Suk kim and Soo-Chang Rey}
\email{star4citizen@gmail.com and screy@cnu.ac.kr}

\author[0000-0003-3474-9047]{Suk Kim}
\affiliation{Department of Astronomy and Space Science \& Research Institute of Natural Sciences, Chungnam National University, Daejeon 34134, Republic of Korea}

\author[0000-0002-0041-6490]{Soo-Chang Rey}
\affiliation{Department of Astronomy and Space Science, Chungnam National University, Daejeon 34134, Republic of Korea}

\author[0000-0002-6261-1531]{Youngdae Lee}
\affil{Department of Astronomy and Space Science \& Research Institute of Natural Sciences, Chungnam National University, Daejeon 34134, Republic of Korea}



\begin{abstract}
The presence of transitional dwarf galaxies in cluster environments supports the hypothesis that infalling star-forming galaxies are transformed into quiescent early-type dwarf galaxies (ETdGs) through environmental effects. We present a study on the evolution of transitional dwarf galaxies, specifically dwarf lenticulars (dS0s) and early-type dwarfs with blue cores (ETdG(bc)s), driven by environmental processes in the Virgo cluster utilizing the Extended Virgo Cluster Catalog. We investigated the morphological fraction and stellar mass of transitional dwarf galaxies in relation to the clustercentric distance, compared to dwarf elliptical galaxies (dEs) and dwarf irregular galaxies (dIrrs). We found that dS0s beyond $0.7R_{\text{vir}}$ exhibit a similar trend in the morphology-clustercentric distance relation to dEs, demonstrating a decreasing fraction with clustercentric distance, whereas ETdG(bc)s display an opposite trend to dS0s but a similar trend to dIrrs. The spatial distributions of transitional dwarf galaxies and dEs correlate with the mass, in which fractions of bright, massive galaxies increase towards the central region of the Virgo cluster. In the mass-clustercentric distance plane, dS0s exhibit a skewed distribution that favors more massive galaxies than dEs at a given clustercentric distance. In the projected phase-space diagram, dS0s are scarce in the stripped region, whereas ETdG(bc)s are absent in both the stripped and virialized regions. In addition, the dS0s in the virialized region are predominantly brighter and more massive than the dEs, indicating that the transformation of dS0s into dEs depends on the stellar mass. We propose that the majority of observed dS0s constitute a population that has settled into the Virgo cluster, whereas ETdG(bc)s represent a recently accreted population. We discuss the impact of ram pressure stripping effects on mass-dependent morphological evolution, as well as the time delay between SF quenching and morphological transformation in dwarf galaxies.

\end{abstract}



\section{Introduction} \label{sec:intro}

It is well established that galaxy properties are influenced by the environment in which galaxies reside. Galaxies in high-density environments, such as galaxy clusters, are primarily composed of early-type galaxies without star formation (SF) activity, while late-type galaxies with high SF dominate in low-density field environments \citep{Dressler1980}. Such a trend between morphology and density holds even within galaxy clusters, where the fraction of galaxies with different morphological types varies depending on the local environment (i.e., local density; \citealp{Binggeli1987}). Various mechanisms have been proposed to explain the different properties of galaxies in clusters (see \citealp{Boselli2006, Boselli2022} for details and references therein). The promising mechanisms involve ram pressure stripping (e.g., \citealp{Gunn1972}) and thermal evaporation (e.g., \citealp{Cowie1977}) of gas by the intracluster medium (ICM), galaxy harassment via high-speed encounters \citep{Moore1996, Moore1998}, and starvation or strangulation via suppression of gas infall by a larger halo \citep{Larson1980}. Consequently, galaxy clusters are ideal places to investigate the role of these environmental effects on the evolution of galaxies in great detail.

Owing to their shallow gravitational potential wells, dwarf galaxies in galaxy clusters are more vulnerable to environmental influences than their more massive counterparts (e.g., \citealp{Boselli2022} and references therein), resulting in the expulsion of substantial gas from the galaxy. The removal of most gas reservoirs facilitates the cessation of SF activity on relatively short timescales, subsequently leading to the transformation into quiescent dwarf galaxies. This environmentally induced quenching predominantly affects dwarf galaxies with masses below $10^{10} M_{\odot}$ in the local universe (e.g., \citealp{Peng2010}). Furthermore, dwarf galaxies constitute numerically dominant low-luminosity galaxies in clusters \citep{Sandage1985, Binggeli1988, Ferguson1994}. Therefore, they serve as ideal probes for examining the various processes governing galaxy evolution in cluster environments.

In the study of environmental effects on the evolution of dwarf galaxies, transitional dwarf galaxies are particularly intriguing objects. These galaxies are believed to have undergone a transformation from star-forming late-type galaxies to quiescent dwarf elliptical galaxies (dEs) in cluster environments. They exhibit characteristics that are intermediate between those of late-type galaxies and dEs in terms of morphology and SF activity. A typical subclass of transitional dwarf galaxies is the dwarf lenticular galaxies (dS0s; \citealp{Sandage1984, Binggeli1991, Binggeli1993, Aguerri2005, Ann2024} and references therein). Morphologically, dS0s possess an overall smooth and regular structure resembling normal dEs but also display extra features such as asymmetric shapes (e.g., bar, lens, and boxyness) and irregular central structures. It has become evident that there is an increase in the complexity and diversity of their properties \citep{Lisker2006a, Lisker2006b, Kim2010, Paudel2010, Janz2012, Toloba2014, Ann2015, Michea2022}. Of particular interest are dS0s hosting substructures with star-forming blue cores \citep{Lisker2006b, Pak2014, Urich2017, Chung2019, Hamraz2019} and hidden spiral arms/disk \citep{Jerjen2000, Barazza2002, DeRijcke2003, Graham2003, Lisker2006a}. Several mechanisms have been proposed to explain the formation of these dS0s in clusters. Among these, the most plausible processes involve ram-pressure stripping, galaxy harassment, and galaxy merging, all of which are postulated to transform the infalling star-forming late-type galaxies into dEs within clusters (\citealp{Chung2019} and references therein).

Dwarf galaxies are affected differently by environmental effects, with the extent of impact varying as a function of their stellar mass \citep{Peng2010, Thomas2010, RomeroGomez2024}. For instance, the SF quenching of dEs induced by environmental effects depends on their stellar mass, with the quenching time being shorter for low-mass dEs ($\sim 10^6 M_{\odot}$) compared to their higher-mass ($\sim 10^8$–$10^9 M_{\odot}$) counterparts \citep{Wheeler2014, Fillingham2015, Wetzel2015}. Ram pressure stripping, one of the multiple processes responsible for SF quenching, is more effective for lower-mass galaxies in a given environment (\citealp{Boselli2022} and references therein). Even when dwarf galaxies share similar local environments within a cluster (i.e., similar locations in spatial distribution), they are expected to exhibit different evolutionary behaviors if their masses differ. Consequently, it may be intriguing to investigate transitional dwarf galaxies depending on their stellar mass, which may provide valuable insights into the transformation and evolution of dwarf galaxies in cluster environments.

To gain a deeper understanding of galaxy evolution influenced by environmental effects in galaxy clusters, it is essential to consider the dynamical behavior of galaxies. In simulations, a three-dimensional phase space diagram, which can provide six parameters (i.e., all three components of both the position and velocity vectors), serves as a valuable tool for studying the infall and orbital histories of galaxies.
Numerous simulations conducted in cluster environments have elucidated different stages of galaxy infall in a phase-space diagram since they first entered the cluster \citep{Mahajan2011, Muzzin2014, Oman2013, Jaffe2015, Oman2016, Rhee2017, Rhee2020}. In observations of cluster galaxies, a projected phase-space (PPS) diagram provides only their projected clustercentric distances and velocities relative to the cluster mean, based on three parameters (i.e., two coordinates and line-of-sight velocity). Observational analysis of the correlation between the locations of galaxies in the projected phase-space (PPS) diagram and their properties also enables the inference of environmental effects within clusters, particularly ram pressure stripping and tidal mass loss (e.g., \citealp{HernandezFernandez2014, Jaffe2015, Yoon2017, Rhee2017}). Therefore, the PPS diagram is a powerful tool for investigating the quenching of infalling galaxies and their transformation within clusters.

Nearby clusters have played a crucial role in elucidating galaxy evolution because of their accessibility for detailed studies, which is not feasible for more distant systems
\citep{Binggeli1985, Ferguson1989, Godwin1983}. 
In this context, the Virgo cluster, which is the nearest rich cluster from the Milky Way at a distance of $\sim$17 Mpc \citep{Mei2007}, serves as an ideal laboratory for detailed investigation of faint, low-mass dwarf galaxies. Given that the Virgo cluster is considered a dynamically young cluster \citep{Aguerri2005}, a physical connection between the cluster and nearby filamentary structures is evident \citep{Tully1982, Kim2016, Castignani2022a, Castignani2022b}. The Virgo filaments predominantly comprise low-mass dwarf galaxies ($<10^9 M_{\odot}$; $\sim$87\% of the total sample, \citealp{Lee2021}). Moreover, a significant fraction ($\sim$58\%) of these dwarf galaxies are classified as star-forming galaxies \citep{Chung2019}. As a result, it is anticipated that dwarf galaxies located in Virgo filaments continue to be accreting into the cluster. Because of the environmental effects within the cluster, the evolutionary pathways of infalling star-forming dwarf galaxies are expected to change, ultimately giving rise to a substantial population of transitional dwarf galaxies within the cluster.

The classical Virgo Cluster Catalog (VCC; \citealp{Binggeli1985}) provides a morphological classification of galaxies in the Virgo Cluster region, based on visual inspection of blue-sensitive photographic plates. However, only a limited number of dS0s have been identified in the VCC \citep{Sandage1984, Binggeli1991}, primarily due to the misclassification of many dS0s in the catalog (see \citealp{Kim2014} for details). An additional limitation is that the VCC predominantly covers the central region within the virial radius ($R_{\text{vir}}$) of the Virgo cluster, thus neglecting outskirts of the cluster. In contrast, the Extended Virgo Cluster Catalog (EVCC; \citealp{Kim2014}) refined the morphological classification and secured a larger sample of dS0s based on a careful examination of SDSS multi-band images. Furthermore, the EVCC encompasses a wider area reaching out to the outskirts of the Virgo cluster (i.e., 3.5 times the $R_{\text{vir}}$ of the Virgo cluster). Thus, the EVCC is a valuable resource for studying the evolution of dwarf galaxies in various environments within the Virgo cluster, including those beyond $R_{\text{vir}}$.

In this study, we aim to investigate the evolution of dwarf galaxies driven by environmental processes in the Virgo cluster by utilizing the extensive galaxy catalog provided by the EVCC. Focusing on transitional dwarf galaxies, we examine their spatial distribution and orbital histories given by the PPS diagram in relation to the stellar mass. We attempt to enhance our understanding of the mechanisms governing galaxy evolution in dense environments, with particular emphasis on how ram pressure stripping affects the morphological evolution of dwarf galaxies. The remainder of this paper is organized as follows. Section 2 describes the dataset and selection of transitional dwarf galaxies in the Virgo cluster. Section 3 presents the projected spatial distribution and morphology-clustercentric distance relation of dwarf galaxies. We also present the stellar mass-clustercentric distance distribution and PPS diagrams of transitional dwarf galaxies in comparison with those of the dEs. Finally, we discuss and summarize our results in Section 4 and Section 5, respectively.

\section{Data and Selection of Transitional Dwarf Galaxies} \label{sec:Data}

Our basic sample of galaxies in the Virgo cluster was taken from the EVCC. We refer readers to \citet{Kim2014} for details on the EVCC data. The EVCC includes photometric and spectroscopic parameters, as well as morphological information of galaxies, which we utilize for our analysis. The EVCC contains 1589 galaxies with radial velocities, 676 of which are absent in the classical VCC \citep{Binggeli1985}. The EVCC covers a rectangular region of 725 deg$^2$ or 60.1 Mpc$^2$, assuming a distance to the Virgo cluster of 16.5 Mpc \citep{Jerjen2004, Mei2007}. This is approximately 5.2 times larger than the area of the VCC and extends to approximately 3.5 times the $R_{\text{vir}}$ of the Virgo cluster. Thus, the EVCC provides an extensive sample of galaxies that spans a wider range of galaxy densities, which is significantly different from the inner region of the Virgo cluster.

Morphological classification of the EVCC was carried out through careful visual inspection of the $g$-, $r$-, and $i$-band combined color images from the SDSS data. These color images offer a higher quality of morphological classification than the VCC, which is based on single-band photographic plates. The vast majority ($\sim$90\%) of galaxies in the EVCC agree with the morphological classification of the VCC. The difference in morphological classification between the two catalogs typically shows the level of one or two morphological subclasses \citep{Kim2014}.

To investigate the morphological transformation of dwarf galaxies, we selected dS0s as our primary sample of transitional dwarf galaxies. According to the morphological classification scheme of the EVCC, dS0s have an overall smooth structure but also exhibit extra features that differ from dEs: a pronounced bulge-disk transition, asymmetric features, and/or a central irregularity (see \citealp{Binggeli1991}, \citealp{Kim2014} for details). The EVCC contains 120 dS0s, which is approximately three times the number of dS0s (44) found in the VCC. Among the EVCC dS0s, 40 are newly listed galaxies that are not included in the VCC. Of the remaining 80 EVCC dS0s, 36 are also identified as dS0s in the VCC. However, it is worth noting that 44 EVCC dS0s are classified as different morphologies in the VCC (see Sec. 6.1 and Figure 15 of \citealp{Kim2014} for details). Most of these galaxies are classified as elliptical (E), lenticular (S0), and dE types in the VCC, although they show distinct bar, lens, and/or central irregularities in the SDSS images (see Figure 19 for examples of \citealp{Kim2014}). To confirm the morphology of galaxies, we examined the Dark Energy Camera Legacy Survey (DECaLS) images of EVCC dS0s that are classified as dEs in the VCC. Figure 1 shows the examples of four VCC dEs reclassified as dS0s in the EVCC. For each galaxy, the image on the right panel displays the residual image provided by DECaLS, which results from the ellipse fit and the subsequent modeling of the light distribution. As shown in residual images, EVCC dS0s exhibit clear substructures (e.g., spiral arm, bar, and disk) hidden by a smooth and symmetric overall light distribution (e.g., \citealp{Lisker2006a}).

The EVCC introduced a complementary classification scheme for galaxies based on their SDSS spectroscopic features (`secondary morphology') in addition to the traditional morphological classification by visual inspection of images (`primary morphology'). The secondary morphology of a galaxy is determined by the shape of its spectral energy distribution (SED) and the presence of H$\alpha$ absorption/emission lines. These characteristics are related to the color and SF activity of galaxies, respectively (see \citealp{Kim2014} for details). The secondary morphology of a galaxy primarily reflects spectral information from its central region owing to the 3 arcsec fiber aperture of the SDSS spectrograph. In our study, we also consider the early-type dwarf galaxies (ETdGs; i.e., dEs and dS0s) that exhibit recent or ongoing SF activity at their centers based on secondary morphology, in addition to the dS0 sample. We selected 122 ETdGs with blue absorption (BA) or blue emission (BE) types from the EVCC. These galaxies are characterized by central stellar populations with an overall SED shape typical of blue galaxies with ages less than 1 Gyr and recent (BA type) or ongoing (BE type) SF activity. Approximately 18\% of ETdGs in the EVCC are of the BA and BE types. Notably, the fraction of these types is significantly larger in dS0s ($\sim$46\%) than in dEs ($\sim$12\%). Because the majority of these galaxies display a blue core at their centers, for subsequent analysis, they are referred to as blue-cored ETdGs (ETdG(bc); see Figure 8 of \citealp{Kim2014} and \citealp{Lisker2006b} for images of these galaxies).

\begin{figure*}[ht!]
\epsscale{1.}
\plotone{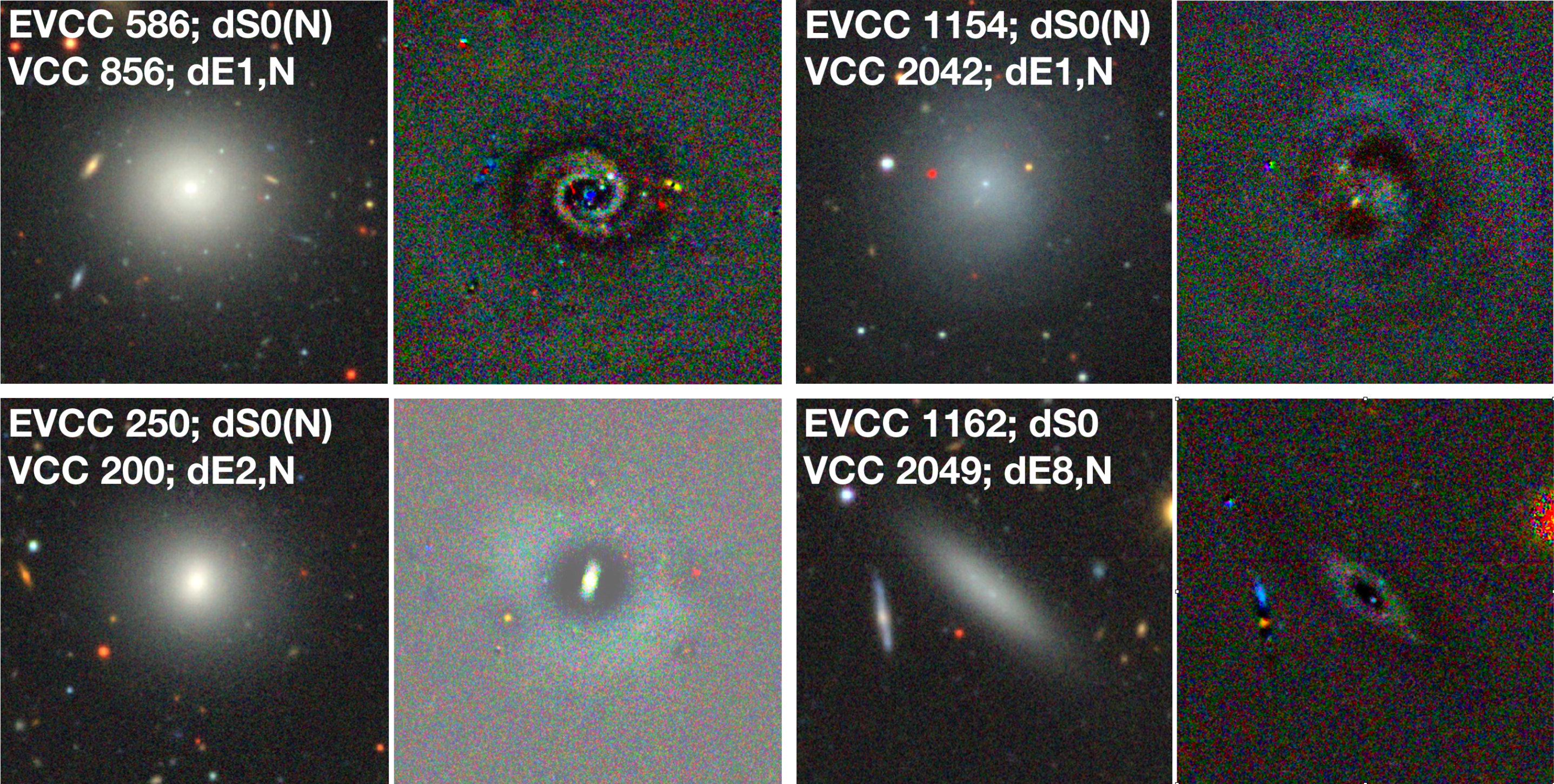}
\caption{
Examples of four VCC dEs reclassified as dS0s in the EVCC. In each galaxy, the left panel denotes $g$-, $r$-, and $i$-band combined color image of DECaLS data, which provides greater depth than the SDSS image. The right panel shows the residual image provided by DECaLS resulting from the ellipse fit and the subsequent modeling of the light distribution. The residual images clearly show unambiguous disk substructures, including spiral arms for EVCC 586 and EVCC 1154, a bar for EVCC 250, and a disk for EVCC 1162.
}
\label{fig1}
\end{figure*}

For our analysis, the ultraviolet (UV) photometric parameters of galaxies are also taken from the NASA-Sloan Atlas (NSA) catalog\footnote{\url{http://nsatlas.org}}, which is based on the Galaxy Evolution Explorer (GALEX) UV data. The stellar masses of galaxies were derived using the relation between the SDSS $g-i$ color and the stellar mass-to-light ratio, derived from the $i$-band luminosity ($L_i$; \citealp{Bell2003}) assuming the initial mass function of \citet{Kroupa1993} and a distance modulus of 31.1 mag corresponding to the Virgo cluster distance \citep{Mei2007}.

\[
\log\left(\frac{M_*}{M_{\odot}}\right) = -0.152 + 0.518(g - i) + \log\left(\frac{L_i}{L_{\odot}}\right)
\]
\\

\section{Results} \label{sec:results}
\subsection{Projected Spatial Distribution of Galaxies} \label{sec: psd}

\begin{figure*}
\includegraphics[width=1.\linewidth]{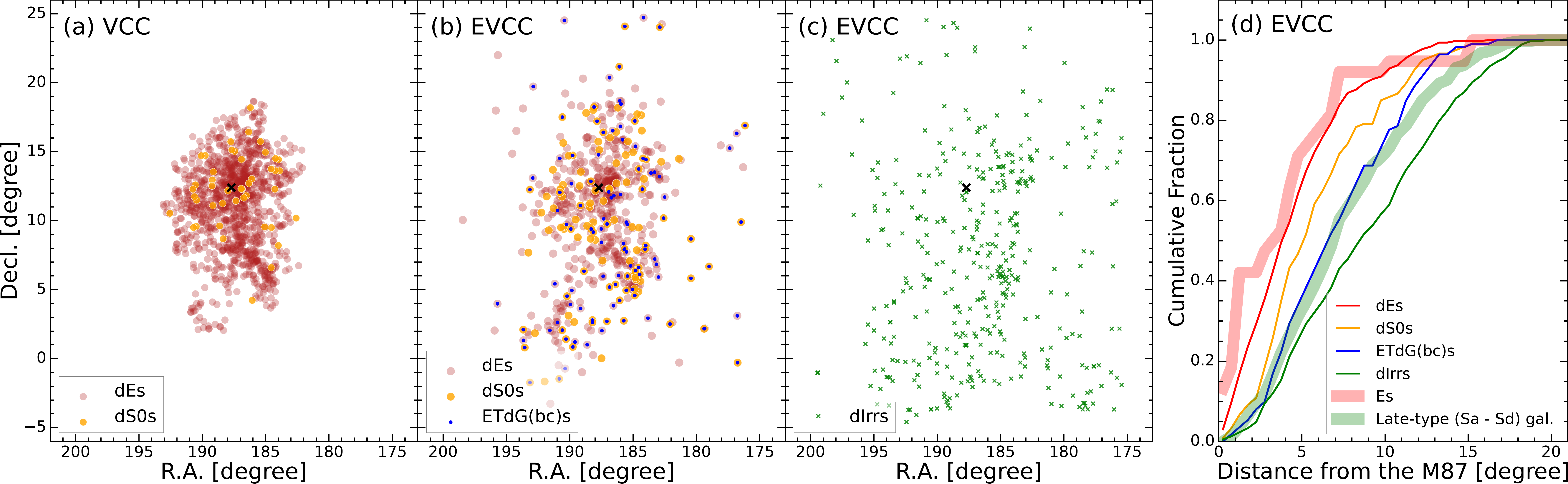}
\caption{(a, b, and c) Projected spatial distributions of galaxies with different morphological types. Panel (a) shows the spatial distribution of dEs (red circles) and dS0s (yellow circles) in the VCC. Panels (b) and (c) show the spatial distributions of dEs (red circles)/dS0s (yellow circles)/ETdG(bc)s (blue dots) and dIrrs (green crosses) in the EVCC, respectively. (d) Cumulative radial distributions of different morphological types in the EVCC as a function of distance from the M87 (black cross in panels a, b, and c). Various line types provide distributions for dEs, dS0s, ETdG(bc)s, dIrrs, Es, and late-type (Sa - Sd) galaxies.
}
\label{fig2}
\end{figure*}

Galaxies with different morphologies are known to have significantly different projected spatial distributions in clusters (e.g., \citealp{Dressler1980}). Figure 2 shows the overall projected spatial distribution of ETdGs, including dEs (red circles), dS0s (yellow circles), and ETdG(bc)s (blue dots), in the VCC (Figure 2a) and EVCC (Figure 2b). We also illustrate the distribution of dwarf irregular galaxies (dIrrs, green circles) drawn from the EVCC for comparison (Figure 2c). First, galaxies in the EVCC are distributed beyond the VCC region. The ETdGs in the EVCC are predominantly located in the central region of the Virgo cluster. In contrast, dIrrs are relatively more dispersed and randomly distributed. The regions outside the VCC that are only covered by the EVCC are mostly dominated by dIrrs. However, a non-negligible fraction of ETdGs in the EVCC is also located on the outskirts of the cluster. In particular, the EVCC covers the ETdGs in the southern extension of the Virgo cluster at Decl. $<$ 5 deg (e.g., \citealp{Vaucouleurs1973}, \citealp{Tully1982}, \citealp{Hoffman1995}, \citealp{Karachentsev2013}), while the VCC contains only ETdGs in the northern part of this structure. The distribution of dS0s in both the EVCC and VCC appears to differ from that of the dEs. While the dEs are strongly concentrated towards the central region of the Virgo cluster, the dS0s exhibit a more scattered distribution. Additionally, ETdG(bc)s also show a spatial distribution that is different from that of the dEs, as they do not show central clustering \citep{Lisker2006b, Lisker2007, Lisker2009, Kim2010, Urich2017}.

For quantitative analysis, we present the cumulative distributions of dEs (red curve), dS0s (yellow curve), and ETdG(bc)s (blue curve) in the EVCC as a function of the distance from the giant elliptical galaxy M87, assuming M87 as the cluster center (Figure 2d). For comparison, we also display the distributions of Es (bold red curve), dIrrs (green curve), and late-type galaxies (i.e., Sa - Sd galaxies, bold green curve), in which the stark difference between their distributions is clearly shown. Es are known to be a relaxed population in the Virgo cluster, as evidenced by the most centrally clustered distribution \citep{Conselice2001}. In contrast, dIrrs show the least central concentration and tend to follow a constant galaxy density profile.

The dEs appear to roughly follow the distribution of Es, but exhibit less central clustering. A two-dimensional Kolmogorov–Smirnov (KS) test on the spatial distributions between Es and dEs yields a p-value of 0.02, which is lower than the standard significance level of 0.05. This indicates rejection of the null hypothesis that the two populations share the same spatial distribution. The spatial distributions of dS0s and ETdG(bc)s differ significantly from that of dEs, with dS0s and ETdG(bc)s lying below the distribution of dEs. A two-dimensional KS test reveals significant differences in the spatial distributions of dS0s and ETdG(bc)s compared to dEs, with p-values of less than 0.001.

Intriguingly, dS0s and ETdG(bc)s, which are subsamples of transitional dwarf galaxies, exhibit different spatial distributions. As shown in Figure 2d, ETdG(bc)s are less concentrated towards the cluster center than the dS0s. A two-dimensional KS test between dS0s and ETdG(bc)s yields a p-value of 0.04, which indicates the rejection of the null hypothesis that their distributions are identical. Furthermore, ETdG(bc)s closely match the distribution of late-type galaxies. A two-dimensional KS test between ETdG(bc)s and late-type galaxies yields a p-value of 0.60, suggesting that the null hypothesis of the same spatial distribution for these two populations cannot be rejected.

\subsection{Morphology-Clustercentric Distance Relation}\label{sec: MCDR}

\begin{figure*}
\epsscale{1}
\centering
\includegraphics[width=0.67\linewidth]{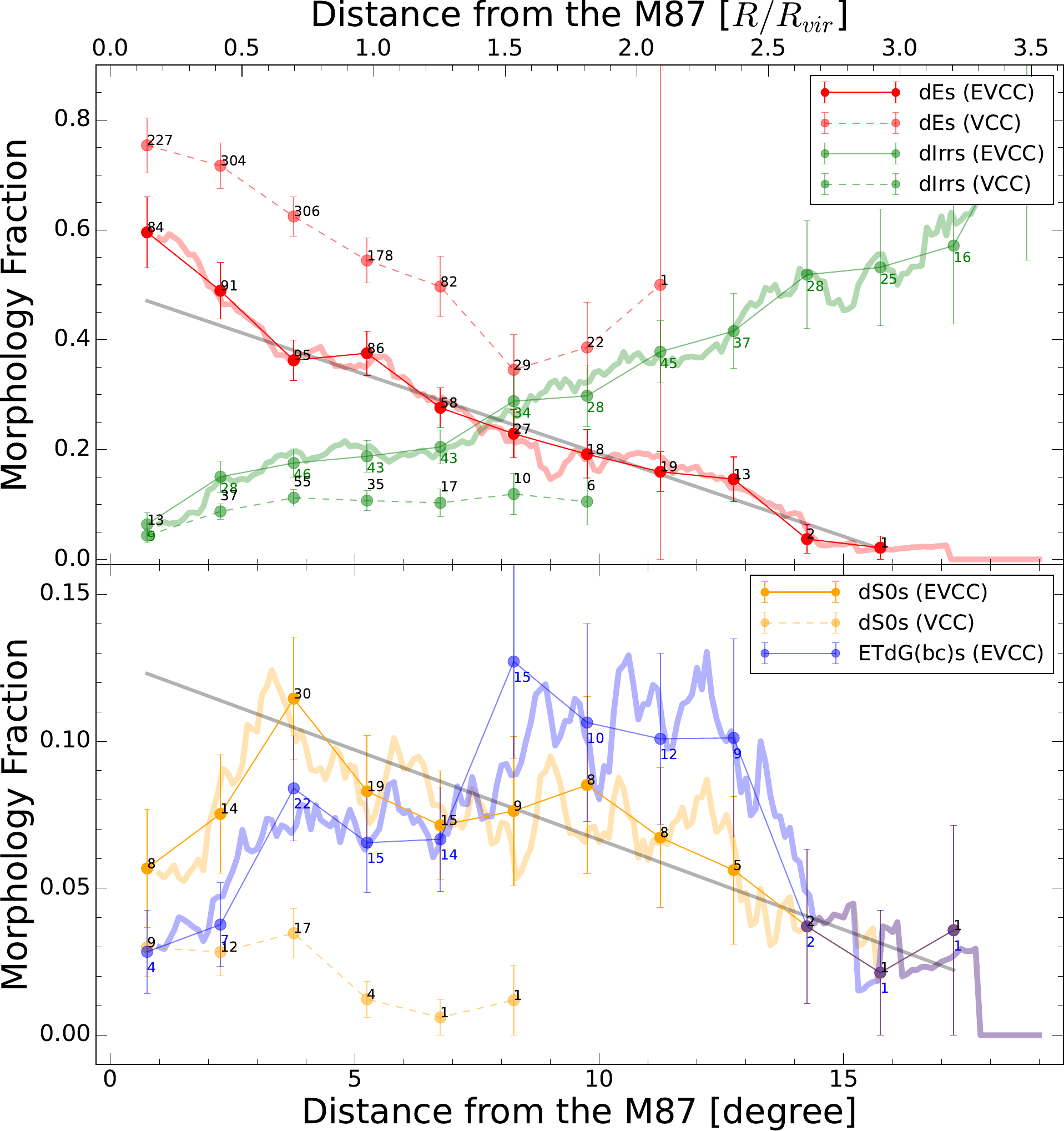}
\caption{Fraction of morphological types as a function of the projected clustercentric distance for dEs and dIrrs (top panel), and dS0s and ETdG(bc)s (bottom panel). The fractions of the different morphological types are shown as filled circles, and the number of galaxies for each filled circle is indicated. The solid and dashed curves indicate the fractions obtained from the EVCC and VCC, respectively. The bold solid curves for the EVCC in both panels are obtained by a running average with a bin width of 2 deg and an incremental step of 0.1 deg. The linear black solid lines are linear least-square fits to the fractions of dEs (top panel) and dS0s (bottom panel) in the outer region beyond 3.5 deg ($\sim$0.7$R_{\text{vir}}$).
}
\label{fig3}
\end{figure*}

In addition to the global trend in the morphology-density relation of massive galaxies \citep{Dressler1980}, the fractions of various types of dwarf galaxies also show a strong correlation with the local density within a given cluster (\citealp{Binggeli1987, Lisker2009} and references therein). It has been proposed that the morphology-clustercentric distance relation is more robust than the morphology-density relation in a cluster environment (e.g., \citealp{Fasano2015}). The morphology-clustercentric distance relation remains largely unchanged regardless of different local densities, whereas the morphology-density relation holds only in the inner regions of clusters \citep{Fasano2015}. Moreover, the morphology-clustercentric distance relation is likely to be more prominent in the low-mass regime \citep{Vulcani2023}.

We derived the fractions of galaxies with different morphological types as a function of projected distance from the M87. In Figure 3, we present the morphology-clustercentric distance relations of dEs and dIrrs (top panel), and dS0s and ETdG(bc)s (bottom panel). The solid and dashed curves indicate the fractions obtained from the EVCC and VCC, respectively. Additionally, the bold solid curves for the EVCC are included in both panels, which are derived through a running average with a bin width of 2 deg and an incremental step of 0.1 deg. We note that the fractions of dEs, dS0s, and ETdG(bc)s located in the outskirts (i.e., beyond about 13 deg or about 2.4$R_{\text{vir}}$) should be interpreted with caution due to the limited number of galaxies. For instance, ETdG(bc)s show an inverted trend in the radial bins of this region. The fractions of all morphological types between the EVCC and VCC differ at all clustercentric distances. This is mainly due to the differences in membership selection and morphology classification between the two catalogs \citep{Kim2014}.

In the top panel of Figure 3, the fractions of dEs in both the EVCC (red solid curves) and VCC (red dashed curve) show a continuous increase towards the cluster center, reaching values of about 60--75\%. However, dIrrs in the EVCC (green solid curves) exhibit an opposite trend to the dEs, displaying a strongly increasing fraction towards the cluster outskirts. In contrast, dIrrs in the VCC (green dashed curve) appear to exhibit a relatively constant trend, without a clear steady increase with the clustercentric distance. The EVCC demonstrates more consistent canonical morphology-clustercentric distance relations for dEs and dIrrs at all clustercentric distances, even in the cluster outskirts beyond $R_{\text{vir}}$.

In the bottom panel of Figure 3, we display the fractions of dS0s (yellow solid curves) and ETdG(bc)s (blue solid curves) in the EVCC, and dS0s (yellow dashed curve) in the VCC. It is worth noting that the dS0s in the EVCC exhibit higher fractions than their VCC counterparts at all clustercentric distances. This discrepancy can be attributed to the reclassification of a large portion ($\sim$55\%, 44 of 80 EVCC dS0s) of non-dS0 dwarf galaxies in the VCC as dS0s in the EVCC (see Section 2 for details). Considering dS0s and ETdG(bc)s as transitional dwarf galaxies evolving towards dEs in the cluster environment, we anticipate that they share comparable morphology-clustercentric distance relations. However, there appear to be different relations between dS0s and ETdG(bc)s. Excluding the radial bins of the central region within 3.5 deg ($\sim$0.7$R_{\text{vir}}$), the fraction of dS0s seems to decrease with increasing clustercentric distance, which is similar to the trend of dEs. In contrast, ETdG(bc)s exhibit an opposite trend to dS0s, showing an increasing fraction with distance if we disregard the uncertain fractions beyond 13 deg. They show a trend similar to that of dIrrs, suggesting that a large fraction of ETdG(bc)s constitutes a recently accreted population in the cluster.

While the dEs follow a well-known morphology-clustercentric distance relation, those in the central region ($<$ 3.5 deg) appear to show a steeper relation than those in the outer region. Furthermore, dS0s in the central region also do not follow their overall trend of increasing fraction with decreasing clustercentric distance, showing a deficiency with a peak in their fraction at around 3.5 deg. We applied linear least-square fits to the fractions of the dEs and dS0s in the outer region beyond 3.5 deg (illustrated by black solid lines in Figure 3). We found that the observed numbers of dEs and dS0s in the central region are about 30 more and 17 less, respectively, than what would be expected based on their relations in the outer region. This suggests that about 60\% of the enhancement of the dEs near the cluster center could be associated with a deficit of the dS0s in the same region. A plausible explanation for this result is that in the central region, where environmental effects are prominent, only bright, massive dS0s are present, whereas the majority of faint, low-mass dS0s have already undergone transformation into dEs (see Sections 3.3 and 3.5 for details).

\subsection{Mass-Clustercentric Distance Distribution}\label{sec: MCD}

\begin{figure*}
\epsscale{1}
\plotone{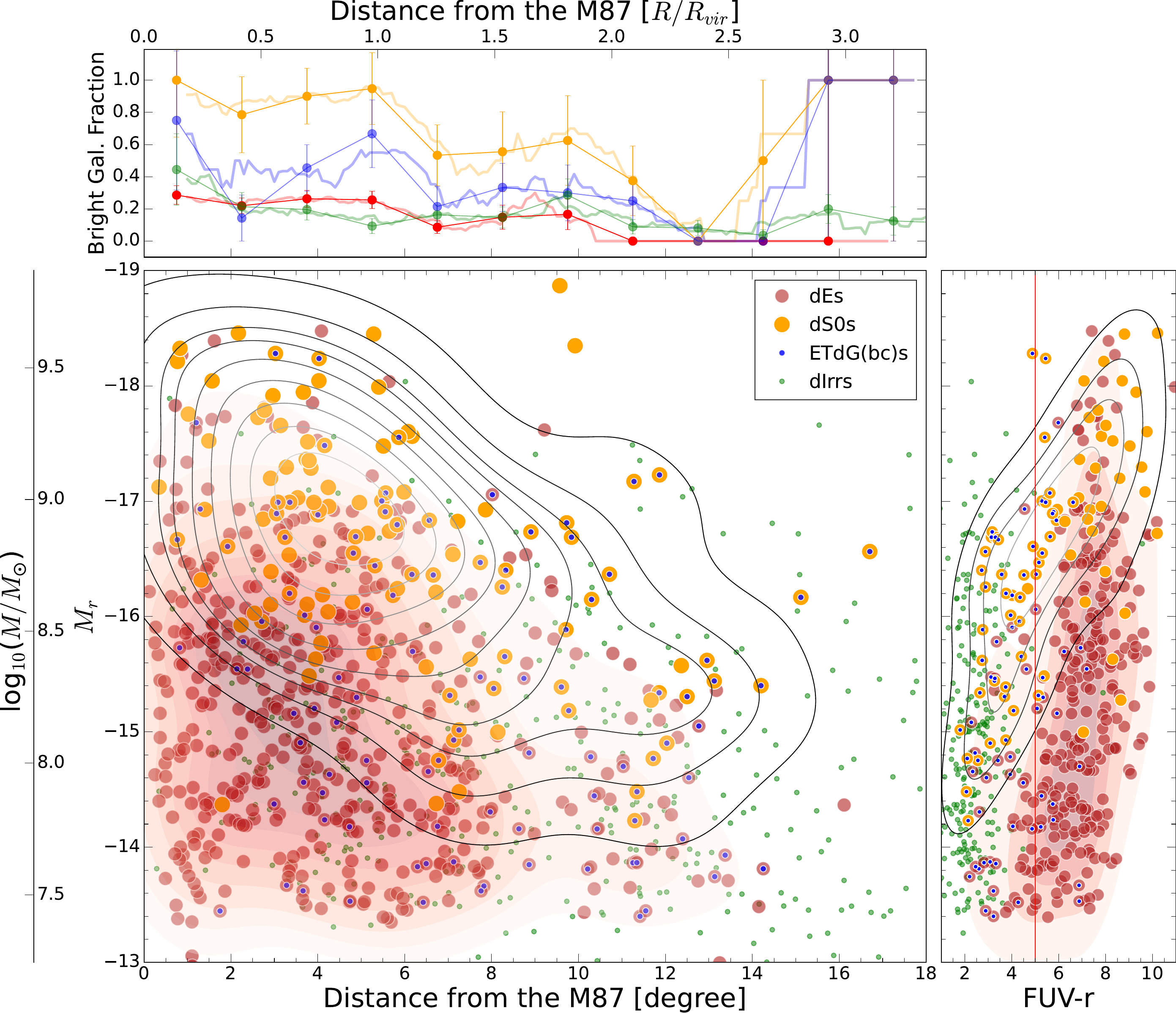}
\caption{(Main panel) Stellar mass (and $r$-band luminosity) versus clustercentric distance for dwarf galaxies with different morphological types in the EVCC: dEs (red circles), dS0s (yellow circles), ETdG(bc)s (blue dots), and dIrrs (green dots). The distributions of dEs and dS0s are highlighted by the red fuzzy and black contours, respectively. 
(Upper panel) Fraction of bright galaxies ($M_r < -16$) against the clustercentric distance for dwarf galaxies of different types: red, yellow, blue, and green circles and lines for dEs, dS0s, ETdG(bc)s, and dIrrs, respectively. The bold solid curves are illustrated for each type of galaxy, obtained using a running average with a bin width of 2 deg and an incremental step of 0.1 deg. (Right panel) FUV-$r$ color-magnitude distribution (CMD) for different types of dwarf galaxies. The symbols and contours are identical to those in the main panel. The red vertical line denotes the color cut of FUV-$r = 5.5$, which is used to differentiate between star-forming and quiescent galaxies.
}
\label{fig4}
\end{figure*}

In Figure 4, we present the distribution of the stellar mass (and $r$-band luminosity) as a function of the clustercentric distance for dwarf galaxies with different morphological types in the EVCC: dEs (red circles), dS0s (yellow circles), ETdG(bc)s (blue dots), and dIrrs (green dots). We also overplot the contours of the number densities for the dEs (red fuzzy contours) and dS0s (black contours) to compare the two subsamples. At first glance, ETdGs, which consist of dEs and dS0s, appear to have a distinct distribution compared to that of ETdG(bc)s and dIrrs. ETdGs exhibit a wedge-shaped distribution, whereas ETdG(bc)s and dIrrs show relatively flat distributions. The ETdGs located in the central region ($<$ 5.4 deg or 1$R_{\text{vir}}$) have a wide range of mass. In particular, massive, bright ETdGs ($> 10^9 M_{\odot}$ and $M_r < -17$) are exclusively found in this region. In contrast, the cluster outskirts are primarily composed of relatively low-mass ETdGs. Another noteworthy feature is that the dEs and dS0s exhibit significantly different distributions. While dEs follow a wedge-shaped distribution, dS0s show a skewed distribution that is biased towards more massive galaxies than dEs at a given clustercentric distance. For instance, in the central region within 1$R_{\text{vir}}$, dS0s predominantly consist of high-mass ($> 10^{8.5} M_{\odot}$) galaxies, whereas the majority of dEs are of relatively lower mass. We note that, in the case of the VCC, such different distributions between dEs and dS0s cannot be clearly visible at cluster outskirts, owing to the paucity of galaxies beyond 8 deg (or 1.5$R_{\text{vir}}$) from the M87 (not shown).

To better understand the mass-clustercentric distance distributions among the various types of dwarf galaxies, in the upper panel of Figure 4, we present the fraction of bright ($M_r < -16$) galaxies for each type: red, yellow, blue, and green circles and lines for dEs, dS0s, ETdG(bc)s, and dIrrs, respectively. Additionally, the bold solid curves are included, which are derived through a running average with a bin width of 2 deg and an incremental step of 0.1 deg. It should be noted again that the fractions of galaxies beyond about 13 deg are highly unreliable owing to the statistically limited sample size of galaxies. The fraction of bright dIrrs appears to be independent of the clustercentric distance and remains constant at approximately 20\%. However, the fractions of bright samples of dEs (red line), dS0s (yellow line), and ETdG(bc)s (blue line) increase as one moves towards the central region of the cluster. It is evident that dS0s consistently exhibit a higher proportion of bright galaxies than dEs and ETdG(bc)s, regardless of the clustercentric distance. Notably, the dS0s located within $R_{\text{vir}}$ are primarily composed of bright galaxies, accounting for approximately 90\% of the total. In contrast, even in the cluster core, the fraction of bright dEs does not exceed 30\% of the total. This verifies the distinct distributions between dEs and dS0s in the mass-clustercentric distance plane.

\subsection{Color-Magnitude Distribution}\label{CMD}

The far-ultraviolet (FUV) flux of a galaxy is particularly sensitive to SF activity within a timescale of roughly 0.1 Gyr \citep{Kennicutt1998, Yi2005, Kaviraj2007}. Thus, FUV-$r$ color can be used as a proxy for the specific SF rate (sSFR) and is useful for differentiating between star-forming and quiescent galaxies. In this context, the FUV-$r$ versus $r$ color-magnitude distribution (CMD) is a particularly efficient tool for tracing SF histories of galaxies. Using galaxies extracted from the VCC, \citet{Kim2010} first revealed a significantly different sequence of Virgo dS0s in the FUV-$r$ CMD compared to the dEs, which was not apparent in previous optical CMDs. They found that dS0s follow a steeper sequence with bluer FUV-$r$ colors than dEs at a given magnitude, indicating that dS0s possess younger stellar populations than dEs.

In the right panel of Figure 4, we present the FUV-$r$ CMD for dwarf galaxies with different morphological types in the EVCC. To separate the star-forming galaxies from the quiescent ones, we applied a color cut of FUV-$r = 5.5$, as indicated by the red vertical line. Following the criterion of \citet{Hammer2012}, we define star-forming galaxies as those with FUV-$r < 5.5$, which corresponds to $\log \text{sSFR (yr}^{-1}) > -10.5$. At all luminosities, dEs (red circles and red fuzzy contours) are dominated by quiescent galaxies with FUV-$r > 5.5$, while only a small fraction of faint ($M_r > -15$) dEs exhibit a slightly blueward offset. In contrast, all dIrrs are located in the blue cloud of the CMD, characterized by blue colors of FUV-$r < 5.5$ (green dots). Our FUV-$r$ CMD reveals that dS0s (yellow circles and black contours) exhibit a distinct distribution separated from dEs, as demonstrated by \citet{Kim2010}. In the luminosity range of $-17 < M_r < -16$, dS0s follow a steeper sequence than dEs, which exhibits a wider FUV-$r$ color distribution extending to bluer colors compared to the mean distribution of dEs. At the brighter end ($M_r < -17$) of the distribution, most dS0s have red colors (FUV-$r > 5.5$) comparable to those of the dEs, indicating quiescent galaxies. At the fainter end ($M_r > -16$) of the distribution, the dS0s show, on average, predominantly blue colors (FUV-$r < 5.5$) and are linked to dIrrs, suggesting that they are star-forming galaxies. A considerable number of faint ($M_r > -17$) dS0s and dEs with FUV-$r < 5.5$ correspond to ETdG(bc)s, which is consistent with their SDSS spectral features indicating signs of recent or ongoing SF activities in their central regions (see Section 2).

In conjunction with the mass-clustercentric distance distribution, our FUV-$r$ CMD demonstrates that the FUV-$r$ colors of dS0s depend on their clustercentric distance. In the central region of the Virgo cluster, bright (i.e., massive) and red (i.e., quiescent) dS0s are prevalent, whereas the majority of faint (i.e., low-mass) and blue (i.e., star-forming) dS0s are found in the outer region of the cluster. These findings are consistent with the results for dwarf galaxies in other clusters, where ETdGs showing signs of SF are preferentially located in the cluster outskirts \citep{Drinkwater2001, Conselice2003, Smith2008, Smith2009}.

\subsection{Projected Phase-space Diagram}\label{PPS}

\begin{figure*}
\epsscale{1}
\plotone{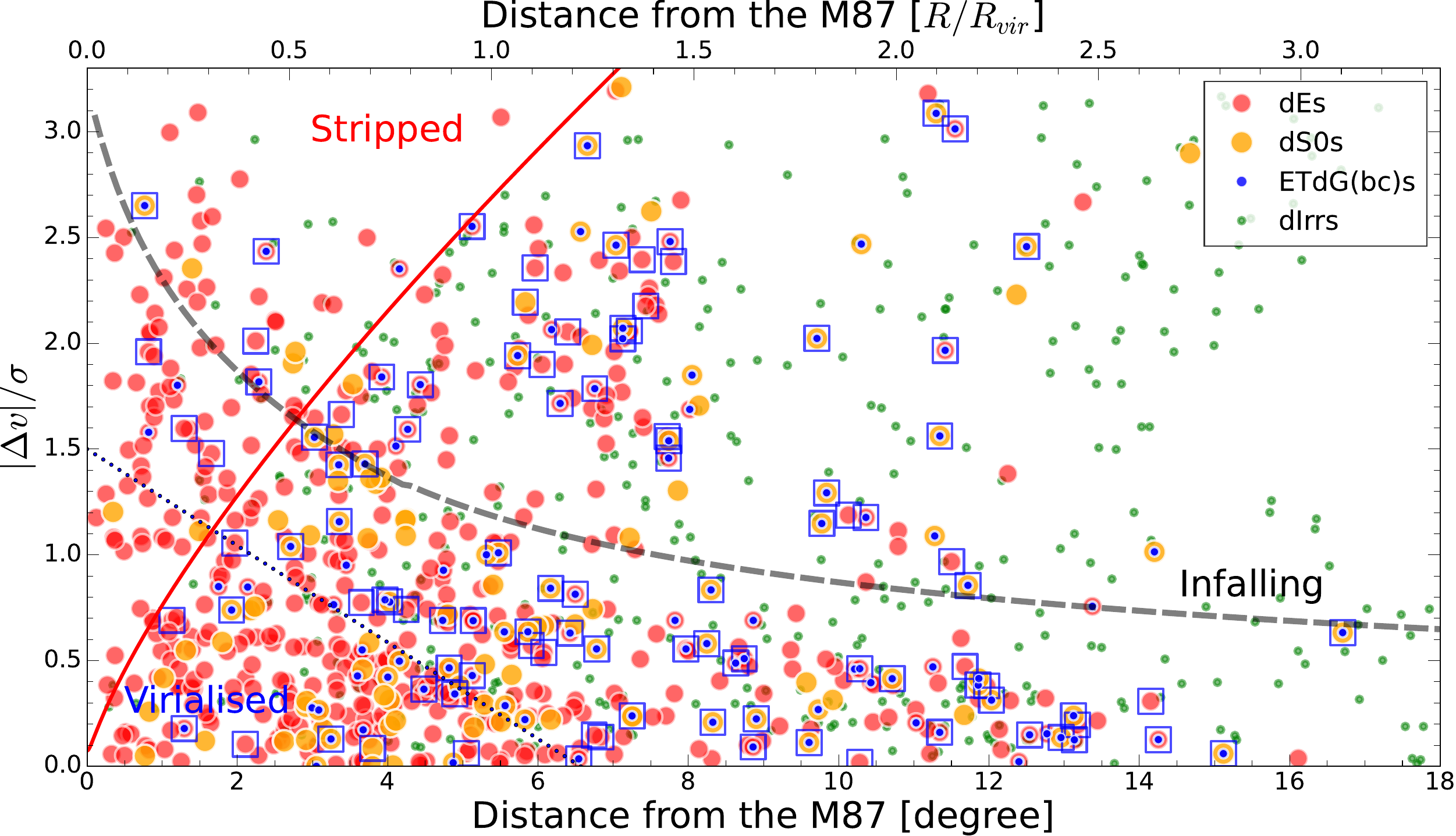}
\caption{PPS diagram of dwarf galaxies with various morphological types in the EVCC: dEs (red circles), dS0s (yellow circles), ETdG(bc)s (blue dots), and dIrrs (green dots). The y-axis is the line-of-sight velocity relative to the cluster mean radial velocity ($\Delta v$), normalized by the velocity dispersion of the Virgo cluster ($\sigma_{\text{cl}}$). Open squares denote ETdGs exhibiting SF with FUV-$r < 5.5$. The black dashed curve indicates the escape velocity of the Virgo cluster. The red solid curve delimits the region where the ram pressure exceeds the restoring force provided by the gravitational potential of a galaxy with a stellar mass of $\sim 10^8 M_{\odot}$. The dotted line represents the virialized region.}
\label{fig5}
\end{figure*}

Figure 5 presents the PPS diagram of the dwarf galaxies in the EVCC. The x-axis represents the projected clustercentric distance from the M87. The y-axis represents the line-of-sight velocity relative to the cluster mean radial velocity ($\Delta v$) normalized by the cluster velocity dispersion $\sigma_{\text{cl}}$. The values of the mean radial velocity $v_{\text{rad}} = 1088$ km s$^{-1}$ \citep{Mei2007}, $R_{\text{vir}} = 1.55$ Mpc \citep{McLaughlin1999, Ferrarese2012}, and $\sigma_{\text{cl}} = 593$ km s$^{-1}$ \citep{Mei2007} of the Virgo cluster are adopted. We include several guidelines in Figure 5 for reference, following the prescriptions presented by \citet{Jaffe2015} and \citet{Yoon2017}. The black dashed curve represents the escape velocity curve for the Virgo cluster, determined by assuming a mass of $4.2 \times 10^{14} M_{\odot}$ that follows an NFW potential model \citep{Navarro1996, McLaughlin1999}. This curve delineates the boundary below which galaxies are gravitationally bound to the cluster. The red solid curve delimits the region where the ram pressure exceeds the anchoring force provided by the gravitational potential of a galaxy with a stellar mass of $\sim 10^8 M_{\odot}$. This curve is derived from the relation $\rho_{\text{ICM}} v_{\text{gal}}^2 > 2 \pi G \Sigma_s \Sigma_g$ \citep{Jaffe2015}, where $\rho_{\text{ICM}}$ denotes the density of the ICM, $v_{\text{gal}}$ represents the velocity of the galaxy relative to the ICM, $G$ is the gravitational constant, and $\Sigma_s$ and $\Sigma_g$ are the surface densities of the stars and gas, respectively. The dotted line demarcates the virialized region where galaxies have likely undergone multiple pericentric passages. This region is approximately defined by $r \lesssim R_{\text{vir}}$ and $\Delta v < 1.5\sigma$ \citep{Mahajan2011}.

We categorize galaxies into three regions in the PPS diagram to examine the effect of ram pressure stripping and the orbital histories of galaxies. The area to the left of the red solid curve represents the region where galaxies are expected to experience complete gas stripping owing to ram pressure, as they attain sufficient velocities after falling into the cluster (the stripped region). The area below the dotted line depicts the region where galaxies have settled into the cluster through virialization (the virialized region). The remaining area outside the stripped and virialized regions comprises galaxies that have recently accreted into the cluster (the recently infalled region).

A non-negligible number of galaxies are situated outside the escape velocity curve in the PPS diagram. Approximately 25\% of ETdGs (i.e., $\sim$23\% for dEs and 27\% for dS0s) are located outside the boundary of the escape velocity. This is mainly because the NFW potential model, which is used to define the boundary of the cluster, assumes a spherically symmetric halo that corresponds to observed clusters of regular shapes. However, the halos of irregular clusters are unlikely to be spherically symmetric. For instance, the Virgo cluster exhibits an irregularly shaped galaxy distribution with some accreting substructures, because it is a dynamically young system that is still in the process of assembly. Thus, the escape velocity curve for the Virgo cluster should be considered as an ideal guideline \citep{Jaffe2015, Yoon2017, Morokuma-Matsui2021}.

A notable result in Figure 5 is that the stripped region is almost devoid of dS0s (approximately only 5\% of the total number of dwarf galaxies) and is largely composed of dEs (76\%). The virialized region, which does not overlap with the stripped region, predominantly consists of dEs (71\%) and dS0s (14\%). In contrast, the recently infalled region is primarily occupied by dIrrs (42\%). However, it is worth noting that a significant proportion of dEs (44\%), dS0s (15\%), and ETdG(bc)s (16\%) is also present in this region. The presence of these ETdGs in the recently infalled region has been explained by a combination of backsplash galaxies, stripped galaxies caused by ram pressure in the cluster outskirts, and preprocessed galaxies in other environments before falling into the cluster (see \citealp{Bahe2013, Jaffe2015} and references therein).

The distribution of UV-detected galaxies in the PPS diagram is an effective tool for tracing SF activity in relation to the orbital histories of galaxies after their infall into a cluster \citep{HernandezFernandez2014, Jaffe2015, Jaffe2016}. In this regard, it is worth noting that Figure 5 displays a notable absence of ETdGs exhibiting SF with FUV-$r < 5.5$ (squares) in both the stripped and virialized regions. This is in line with previous findings of a lack of HI-detected galaxies in the stripped and virialized regions of the A963\_1 cluster \citep{Jaffe2015} and the Virgo cluster \citep{Yoon2017}, as the SFR of galaxies is correlated with their HI content. Our results support the scenario of quenching of SF by ram pressure stripping and subsequent morphological transformation of dwarf galaxies depending on their orbital histories. When late-type, star-forming galaxies enter a massive cluster from the field and filamentary structure (i.e., the recently infalled region), they undergo HI gas stripping during their first passage through the ram pressure stripping region. The SF of galaxies that have most likely experienced several pericentric passages is reduced and eventually quenched once the HI gas is almost completely removed (i.e., the virialized region).

\begin{figure*}
\epsscale{1}
\plotone{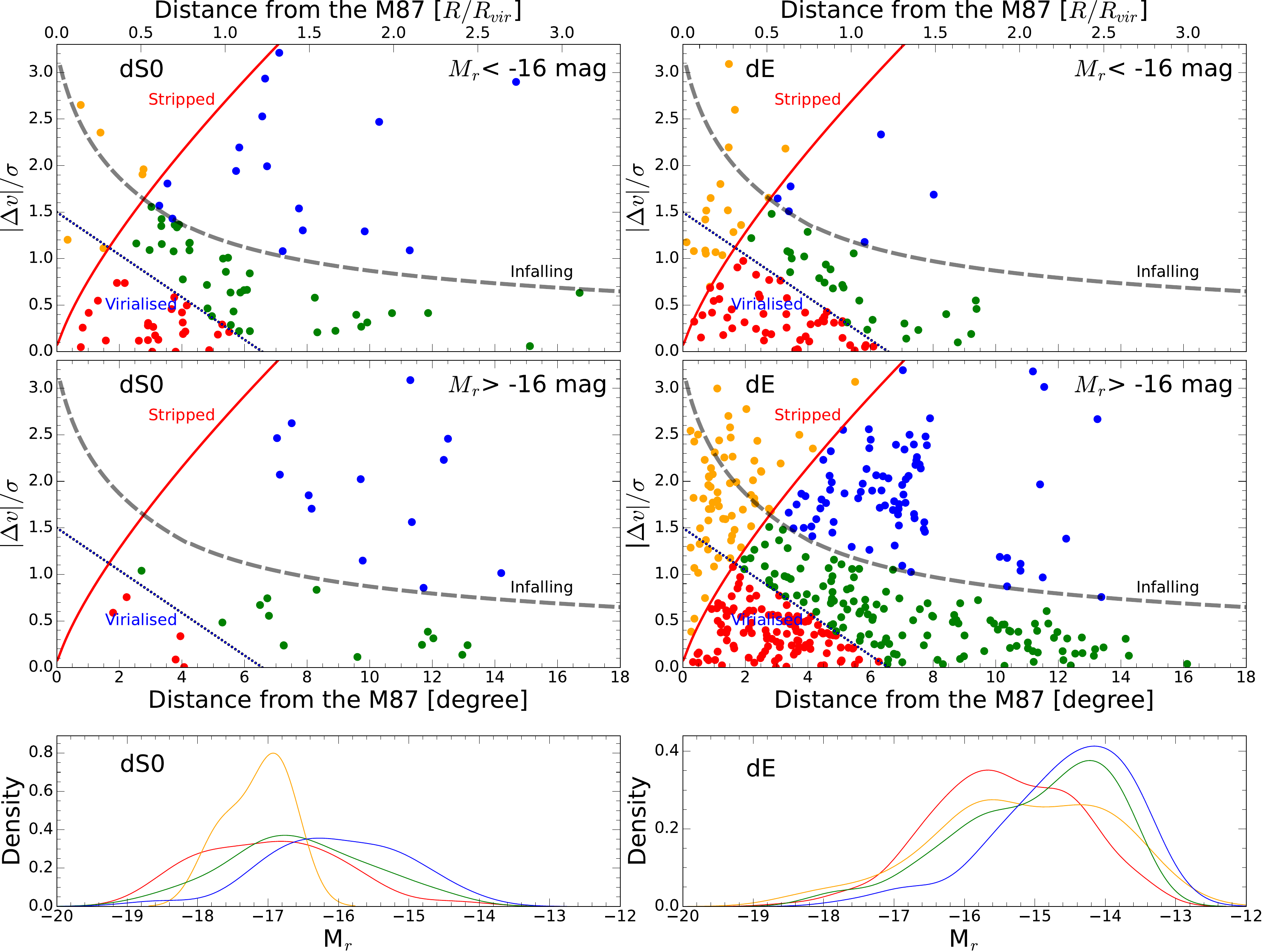}
\caption{(Top and middle panels) PPS diagrams of bright ($M_r < -16$, top panels) and faint ($M_r > -16$, middle panels) dEs (left panels) and dS0s (right panels). Filled circles with various colors represent galaxies in different regions: yellow for the stripped region, red for the virialized region, green for the recently infalled region, and blue for outside the escape velocity curve. All curves are identical to those shown in Figure 5. (Bottom panels) Luminosity distributions of dS0s (left panel) and dEs (right panel) in each region of the PPS diagram. The colors of the curves denote different regions and correspond to the colors of the galaxies shown in both the upper and middle panels.}
\label{fig6}
\end{figure*}

As shown in Figure 5, the dEs are abundant across all regions in the PPS diagram. However, dS0s are not common in the stripped region, but are found in the recently infalled and virialized regions. Assuming that dS0s are transitional objects in the process of evolving into dEs due to environmental effects in the cluster, the paucity of dS0s in the stripped region suggests that they have already been transformed into dEs through ram pressure stripping. In this case, it is not plausible to postulate the existence of dS0s in the virialized region at the terminal stage of their orbits within the cluster. Thus, in the context of ram pressure stripping effects linked with the orbital histories of galaxies in the cluster (e.g., \citealp{Jaffe2015, Yoon2017}), the scarcity of dS0s only in the stripped region (i.e., first pass through the core of dense ICM) may be a particularly intriguing result.

As demonstrated in Figure 4, considering the result that the stellar mass of dS0s varies with clustercentric distance, it is interesting to see how the distributions of the different galaxy populations differ in the PPS diagram, based on their luminosity (i.e., stellar mass). In Figure 6, we present PPS diagrams for bright, massive ($M_r < -16$ or $> 10^{8.6} M_{\odot}$, top panels) and faint, less massive ($M_r > -16$ or $< 10^{8.6} M_{\odot}$; middle panels) dS0s (left panels) and dEs (right panels). It is evident that the dS0s in the virialized region are dominated by bright, massive ones, whereas faint dS0s are lacking. Another intriguing feature is that faint dEs are abundant in the stripped region, whereas faint dS0s are absent in this region. In the bottom panels of Figure 6, we present the luminosity distributions for the dS0s (left panel) and dEs (right panel) within each region of the PPS diagram. We confirm that both luminosity distributions for dS0s in the stripped (yellow curve; median value of $M_r = -17.1$) and virialized regions (red curve; median value of $M_r = -16.9$) are systematically offset towards brighter luminosities compared to dEs in the same region (median values of $M_r = -15.2$ and $M_r = -15.5$ for the stripped and virialized regions, respectively).

\section{Discussion}\label{discussion}

Several studies have extensively discussed the phenomenon of ram pressure stripping for dwarf galaxies. Ram pressure stripping operates more effectively on lower-mass galaxies in a specific environment (e.g., \citealp{Murakami1999}, \citealp{Mori2000}, \citealp{Marcolini2003}, \citealp{Bekki2009, Bekki2014}, \citealp{Wheeler2014}, \citealp{Emerick2016}, \citealp{Fillingham2016}, \citealp{Boselli2022}, \citealp{Greene2023}). In particular, \citet{Mori2000} estimated the gas-removal timescale for dwarf galaxies in clusters caused by ram pressure stripping as a function of mass and clustercentric distance of dwarf galaxies. At a given clustercentric distance, low-mass dwarf galaxies with mass less than $10^9 M_{\odot}$ completely lose their gas within 1 Gyr, whereas more massive dwarfs can retain gas for longer timescales (see Figure 5 of \citealp{Mori2000}).

If ram pressure plays a significant role in the quenching and evolution of galaxies in cluster environments, the mass-dependent effect of ram pressure stripping is consistent with our results that the transformation of dS0s into dEs depends on their stellar mass. Bright and massive dS0s are more resistant to the ram pressure stripping effect owing to their deeper gravitational potential wells, making it more difficult to change their morphology into dEs. As depicted in Figure 4, dS0s are systematically more massive than dEs at a given clustercentric distance, indicating that massive dS0s have not yet been transformed into dEs. Furthermore, massive dS0s are more commonly found near the cluster center as a result of dynamical friction (Figures 4 and 6). These massive dS0s can maintain their morphology by the time they settle into the virialized region (more than 4 Gyr since cluster infall, \citealp{Jaffe2015}). However, faint and low-mass dS0s can be rapidly transformed into faint dEs upon entering the cluster owing to their sensitivity to the ram pressure effect. This is responsible for the scarcity of faint dS0s and the abundance of faint dEs across all regions in the PPS diagram. This also explains the distinct distribution of dS0s in the mass-clustercentric distance diagram, where only bright, massive dS0s are found in the central region of the cluster.

dS0s with morphological substructures, such as disk, spiral arm, and bar, have been identified in the Virgo cluster \citep{Jerjen2000, Barazza2002, Lisker2006a}. A systematic search for dS0s with disk substructures in the Virgo cluster revealed that the fraction of such galaxies is approximately 25\% at $M_r < -16$ and even exceeds 50\% at the bright end of the ETdGs \citep{Lisker2006a, Lisker2009}. Furthermore, kinematic studies of dS0s in the Virgo cluster show that these galaxies are not pressure-supported systems; instead, they are sustained by fast rotation \citep{Toloba2011, Toloba2015}. As evident from Figure 6, many bright dS0s are located in the virialized region, indicating that they have been in the cluster for a long time since their infall. This suggests that the extra substructures in their morphologies and internal kinematics have yet to be completely destroyed. However, as evident from the CMD in Figure 4, these bright dS0s are quenched, with old ages inferred from their FUV-$r$ colors ($> 5.5$) comparable to those of the dEs. In contrast, for faint and lower-mass ETdGs, quenched dS0s are rare, whereas the majority of quenched galaxies are dEs. This implies that the SF quenching of low-mass dEs is accompanied by morphological changes.

A plausible explanation is that the SF of bright, massive dS0s is suppressed earlier than their structural transformation into dEs owing to cluster environmental effects. Despite the cessation of SF, the substructures of the massive dS0s may persist for some time. Indeed, there is growing evidence that the timescale for the morphological transformation of galaxies differs from that of SF quenching (e.g., \citealp{Bamford2009}, \citealp{Wolf2009}, \citealp{Jaffe2011}, \citealp{Cortese2019}, \citealp{Kelkar2019}, \citealp{Martinez2023}). For instance, galaxies in clusters, such as red spiral galaxies and satellites around a host galaxy, undergo delayed structural changes compared with their quenching phase. In particular, morphological transformation is a slower process that occurs on timescales greater than the cluster crossing time ($\sim$ a few Gyr, \citealp{Jaffe2015, Kelkar2019}).

This further suggests that ram pressure stripping, which directly influences the gas content in galaxies, may be the dominant physical process. Because of their shallow gravitational potential wells, most gas reservoirs of dwarf galaxies located in or near the cluster core can be efficiently removed by ram pressure stripping on relatively short timescales ($\sim$150 Myr) compared to the typical crossing time within a cluster, including the Virgo cluster ($\sim$2 Gyr; \citealp{Boselli2006, Boselli2022}). This can lead to subsequent rapid cessation of SF activity \citep{Boselli2008}. However, this process does not significantly affect the distribution and kinematics of the stellar population (i.e., the structure), thereby avoiding substantial morphological disturbances of galaxies.

Recent studies have found that ETdG(bc)s are ubiquitously found in a variety of environments, including clusters \citep{DeRijcke2003, DeRijcke2013, Lisker2006b, Lisker2007, Urich2017, Hamraz2019}, groups \citep{Cellone2005, Tully2008, Pak2014}, filaments \citep{Chung2021, Chung2023}, and the field \citep{Gu2006, Rey2023}. The formation of ETdG(bc)s in high- and moderate-density environments, such as clusters, groups, and filaments, is thought to result from several environmental effects. The most plausible mechanisms are ram-pressure stripping and galaxy harassment for clusters \citep{Lisker2006b, Lisker2007, Lisker2009} and galaxy merging/interaction for groups and filaments \citep{Pak2014, Chung2023}, all of which are responsible for the transformation of late-type galaxies into dEs. It is important to note that ETdG(bc)s are not expected to form in low-density environments, where environmental effects are minimal. However, a significant number of ETdG(bc)s has recently been discovered in isolated environments \citep{Rey2023}. These isolated ETdG(bc)s may be quiescent blue compact galaxies (QBCDs) in the recurrent evolutionary sequence of the BCD and QBCD phases lasting for a few Hubble times (e.g., \citealp{SanchezAlmeida2008}; see \citealp{Rey2023} for details).

In this study, we identified 122 ETdG(bc)s in the Virgo cluster, characterized by their central young stellar populations formed from recent or ongoing SF activity. ETdG(bc)s show a less centrally concentrated spatial distribution, following the distributions of late-type and dIrr galaxies in the Virgo cluster (Figures 2, 3, and 4), whereas dEs dominate the high-density region of the cluster. As depicted in the PPS (Figure 5), most Virgo ETdG(bc)s are found in the infalled region, whereas they are scarce in the stripped and virialized regions. This suggests that the ETdG(bc)s in the Virgo cluster represent a recently infalled and unrelaxed population that has not yet become passive dEs.

It has been reported that the fraction of ETdG(bc)s in moderate- and low-density environments is significantly higher than in clusters. Although ETdG(bc)s in the Virgo cluster comprise only about 20\% of all ETdGs, the majority of ETdGs in groups ($\sim$70\%; \citealp{Pak2014}) and the field ($\sim$91\%; \citealp{Rey2023}) are a population of ETdG(bc)s. This finding provides observational evidence that ETdG(bc)s are more likely to form and survive in low-density environments compared to clusters. Furthermore, the presence of ETdG(bc)s in group, filament, and field environments may be strongly related to the preprocessing that occurs in the structures located outside the cluster. \citet{Chung2021} found evidence of preprocessing for a sample of ETdG(bc)s in filamentary structures surrounding the Virgo cluster \citep{Kim2016, Castignani2022a, Castignani2022b}, wherein ETdG(bc)s in the filament exhibit chemical properties and SF activities comparable to their counterparts in the Virgo cluster. This suggests that the transformation of infalling late-type galaxies by environmental effects within the cluster may not be the primary process in the formation of Virgo ETdG(bc)s. Therefore, a considerable fraction of ETdG(bc)s found in the Virgo cluster, which are mostly located in the infalled region, may have originally been ETdG(bc)s formed in other lower-density environments and subsequently fell into the cluster. Once ETdG(bc)s enter the stripped and virialized regions of the Virgo cluster, they are likely to evolve into passive dEs.

\section{Summary and Conclusions}\label{summary}

In this study, we investigated the morphological evolution of dwarf galaxies in the Virgo cluster, based on their spatial distributions and orbital histories. By utilizing the extensive EVCC catalog data, which extends to approximately 3.5 times the $R_{\text{vir}}$ of the Virgo cluster, we focused on the transitional dwarf galaxies, specifically dS0s and ETdG(bc)s, in various environments within the cluster.  

\begin{enumerate}
    \item We found that dS0s and ETdG(bc)s have less central clustering in their spatial distributions than dEs. Additionally, ETdG(bc)s show less concentration towards the cluster center compared to dS0s and closely follow the spatial distribution of late-type galaxies, suggesting that the ETdG(bc)s represent a recently infalled population.

    \item The EVCC demonstrates stronger morphology-clustercentric distance relations for dEs and dIrrs compared to the classical VCC catalog, even in the cluster outskirts beyond $R_{\text{vir}}$. The dE fraction continuously increases towards the cluster center, whereas the dIrrs exhibit the opposite trend to the dEs, displaying an increasing fraction towards the cluster outskirts. It appears that there are different morphology-clustercentric distance relations between dS0s and ETdG(bc)s. The fraction of dS0s outside 3.5 deg ($\sim$0.7$R_{\text{vir}}$) from the cluster center seems to decrease with increasing clustercentric distance, similar to the trend of dEs. In contrast, ETdG(bc)s exhibit an opposite trend to dS0s, displaying an increasing fraction with clustercentric distance. The lack of dS0s in the central region is attributed to an increase of dEs in the same region, which may be due to the rapid transformation of faint, low-mass dS0s into dEs induced by environmental effects.

    \item In the mass-clustercentric distance plane, the dS0s and dEs exhibit distinct distributions that are not easily observable in the VCC. The dEs in the central region ($<$ 5.4 deg or 1$R_{\text{vir}}$) exhibit a wide range of mass, whereas those in the cluster outskirts are primarily composed of relatively low-mass galaxies. In contrast, dS0s exhibit a skewed distribution that favors more massive galaxies than dEs at a given clustercentric distance. The fractions of bright and massive samples for dEs, dS0s, and ETdG(bc)s increase as one moves towards the central region of the Virgo cluster. Notably, the dS0s located within $R_{\text{vir}}$ are primarily composed of massive ($> 10^{8.5} M_{\odot}$) galaxies.

    \item We confirmed that the dS0s display a unique distribution that is distinct from the dEs in the FUV-$r$ CMD. The dS0s exhibit a steeper sequence than the dEs, with bluer FUV-$r$ colors at a given magnitude. Additionally, the CMD indicates that the FUV-$r$ colors of the dS0s depend on their distance from the cluster center. In the central region of the Virgo cluster, there is a prevalence of massive, quiescent dS0s with FUV-$r > 5.5$, while the majority of low-mass, star-forming dS0s with FUV-$r < 5.5$ is found in the cluster’s outer region.

    \item We compared the distributions of various types of dwarf galaxies in the PPS diagram. The dEs are prevalent in all regions of the PPS diagram. Notably, while a substantial fraction of dS0s is present in the infalled and virialized regions, they are scarce in the stripped region. Additionally, there is a notable absence of ETdGs exhibiting SF with FUV-$r < 5.5$ as well as ETdG(bc)s in both the stripped and virialized regions. This supports the hypothesis that SF quenching of dwarf galaxies by ram pressure stripping, followed by their subsequent transformation, is linked to the infall and orbital histories of galaxies.

    \item We examined the distributions of dwarf galaxies in the PPS diagram based on their luminosity (i.e., stellar mass). We found that dS0s in the virialized region are dominated by brighter and more massive galaxies than dEs in the same region. This suggests that the transformation of dS0s into dEs is dependent on their stellar mass; massive dS0s can maintain their morphology until they reach the virialized region, whereas low-mass dS0s can quickly transform into faint dEs upon entering the cluster due to their sensitivity to environmental effects.
\end{enumerate}

A central finding of this study is the differential evolution of dS0s within the Virgo cluster contingent upon their stellar mass. Low-mass dS0s exhibit high susceptibility to environmental effects, resulting in their rapid evolution into dEs. In contrast, higher-mass dS0s can retain their morphology owing to their substantial gravitational potential, even in the cluster center where environmental influences are intense. This mass-dependent transformation pathway of dwarf galaxies enhances our understanding of the intricate processes involved in galaxy evolution in dense cluster environments. To further investigate the physical mechanisms of galaxy transformations caused by cluster environments, a comparative study of transitional dwarf galaxies across various clusters with different densities and dynamical conditions is required. The presentation of our results on these topics is deferred to forthcoming publications.

\begin{acknowledgments}
We are grateful to the anonymous referee for helpful comments
and suggestions that improved the clarity and quality of this paper. S.K. and S.C.R. contributed equally to this work as corresponding authors. This work was supported by the National Research Foundation of Korea through grants, NRF-2019R1I1A1A01061237 and NRF-2022R1C1C2005539 (S.K.), NRF-2022R1A2C1007721 (S.C.R.), and NRF-2022R1I1A1A01054555 (Y.L).
\end{acknowledgments}

\bibliography{sample631}{}
\bibliographystyle{aasjournal}



\end{document}